\documentclass[ aps,floatfix,  twocolumn, reprint, showpacs, footnoteinbib]{revtex4-1}

\usepackage{graphicx}
\usepackage{amssymb}   % for math
\usepackage{amsfonts}
\usepackage{amsmath}
\usepackage{dsfont}
\usepackage{dcolumn}   % needed for some tables
\usepackage{bm}        % for math
\usepackage[mathscr]{eucal}
\usepackage[dvipsnames]{xcolor}
\usepackage[colorlinks, linkcolor=blue,citecolor=blue,urlcolor=blue]{hyperref}
\usepackage[all]{hypcap} % let hyperlinks correctly point to figures rather than their captions; 
\usepackage{wasysym}

\newcommand{\Le}{\left}
\newcommand{\Ri}{\right}
\newcommand{\nn}{\nonumber}
\newcommand{\f}{\frac}

\newcommand{\mc}{\mathcal}

\newcommand{\ra}{\rangle}
\newcommand{\la}{\langle}

\newcommand{\eq}[1]{\begin{align}#1\end{align}}

\newcommand{\msr}{\mathscr}
\newcommand{\ez}{E=0}
\begin{document}
\title{Generalized Dyson model: nature of zero mode and its implication in dynamics}
\author{Giuseppe De Tomasi}
\author{Sthitadhi Roy}
\author{Soumya Bera}
\affiliation{Max-Planck-Institut f\"ur Physik komplexer Systeme, N\"othnitzer Stra{\ss}e 38,  01187-Dresden, Germany}
% 
% \date{\today}
%
\begin{abstract}
We study the role of the anomalous $E=0$ state in dynamical properties of non-interacting fermionic chains with chiral 
symmetry and correlated bond disorder in one dimension.
These models possess a diverging density of states at zero energy leading to a divergent localization 
length at the band center. By analytically calculating the localization length for a finite system, we show
that correlations in the disorder modify the spatial decay of the $E = 0$ state from being quasilocalized
to extended.
We numerically simulate charge and entanglement propagation and provide evidence that states close to 
$E=0$ dominate the dynamical properties. Remarkably, we find that correlations lead to subdiffusive charge 
propagation, whereas the growth of entanglement is logarithmically slow. A logarithmic scaling  of entanglement saturation 
with  system size is also observed, which indicates a behavior akin to quantum critical glasses. 
\end{abstract}
\maketitle

%%%%%%%%%%%%%%%%%%%%%%%%%%%%%%%%%%%%%%%%%%%%%%%%%%%%%%%%%
% {\it Introduction.}
\section{Introduction \label{sec:intro}}
%%%%%%%%%%%%%%%%%%%%%%%%%%%%%%%%%%%%%%%%%%%%%%%%%%%%%%%%%
In one dimensional disordered systems, the presence of a chiral symmetry~(sub-lattice) can lead to 
some energy eigenstates behaving differently than all other localized eigenstates. For instance, a bond 
disordered model, which we refer to as the Dyson I model~\cite{Dyson53, Weiss75},  has a diverging density of states 
at $\ez$, which is accompanied with a divergent localization length~\cite{Bush75,Th76,Egg78}. 
Although the localization length is diverging, the state is quasilocalized because the
localization length scales sub-extensively with system size~\cite{Lic77, Eco81}. 
% Recently Anderson localization with off-diagonal disorder has been experimentally tested using optical waveguides.
Another mechanism in disordered systems that can modify the nature of its eigenstates is the 
presence of correlations in disorder. 
Even with onsite disorder, where all single particle eigenstates are exponentially localized in one 
dimension~\cite{Mott61,Past}, correlations in the disorder can either 
partially or completely destroy localization~\cite{Dunlap90, Lyra98, Izrailev99, Shima04, Khum08, 
Croy11, Izrailev12, Che05}. Moreover, the study of correlated disorder has several practical applications, 
particularly in transport properties of disordered conducting polymers and biological molecules~\cite{Klotsa05,Krokhin09}.

The combination of symmetries and disorder correlations can have interesting effects in the physics of 
Anderson localization. For instance, bond-dimerization (referred to as Dyson II model~\cite{ZimanPRL82} hereafter), 
{\textit{i.e.}} random bonds appearing in identical pairs,
($J_{2l-1}{=}J_{2l}$, $J_l$ is the bond strength.)
changes the nature of the $\ez$ state from being quasilocalized to extended. Despite this, the role of 
 local disorder correlations with regard to the nature of the $\ez$ state and the consequent effect on non-equilibrium dynamical properties has not been explored 
extensively so far. In this article, we construct and study a random bond model with tunable correlated bond disorder, such that  the 
spatial extension of the $\ez$ state can be modified almost continuously from being exponentially localized to 
extended. The construction also allows us to recover the known Dyson I and II models in appropriate limits.
We further examine the effect of the nature of the $\ez$ state on the transport properties via charge and 
entanglement propagation.

Recently, dynamical properties of isolated 
disordered systems have attracted much attention due to advancement of controlled experimental techniques as well as 
the discovery of dynamical quantum phase transitions.  In particular, dynamical properties are used to 
characterize different localized phases. For example, in both the Anderson localized and the many-body localized~(MBL) 
phase~\cite{baa06, Mirlin05} charge transport is absent. However, while in the 
former the bipartite entanglement $\msr{S}(t)$ does not grow with time, in the latter it grows 
logarithmically~\cite{Bardarson12, AbaG13}.
Furthermore, it has been shown that, while in the ergodic phase of the MBL system charge and entanglement show 
subdiffusive and subballistic behavior respectively~\cite{barlev15,LuitzPRB16}, in a diffusive non-integrable spin-chain  $\msr{S}(t)$ grows ballistically with time~\cite{Kim13}.
It is then natural to conclude that charge and entanglement propagation can have different dynamical behaviors, which 
further motivates us to contrast them in the presence of both 
disorder-correlation and symmetries.
Interestingly, the generalized Dyson model shows subdiffusive density propagation and logarithmic
entanglement growth, a phenomenon that has not been observed previously in disordered systems.

The rest of the paper is organized as follows. In the Sec.~\ref{sec:model}, we introduce the generalized model and analytically derive the localization length of the $E=0$ state, and describe the phase diagram with regard to the localization properties. We describe the dynamical properties in Sec.~\ref{sec:dynamical}, with Sec.~\ref{sec:dyson2} containing the numerical results for the Dyson II limit while the results for other parameter values are presented in Sec.~\ref{sec:dyson3}. Finally, the results are summarized in Sec.~\ref{sec:conclusion}.
%
%%%%%%%%%%%%%%%%%%%%%%%%%%%%%%%%%%%%%%%%%%%%%%%%%%%%%%%%%%%%%%
\section{Model and Localization length \label{sec:model}}
%%%%%%%%%%%%%%%%%%%%%%%%%%%%%%%%%%%%%%%%%%%%%%%%%%%%%%%%%%%%%%
The nearest-neighbor random hopping model is defined as, 
\eq{
\mathcal{H} = -\sum_l [J_{2l-1}c_{2l-1}^\dagger c_{2l} + J_{2l}c_{2l}^\dagger c_{2l+1} + \mathrm{h.c}],
\label{eq:genericham}
}
where $c_l^\dagger~(c_l)$ is the fermionic creation (annihilation) operator at
site $l$ and $J_l$s are positive random hopping amplitudes. 
The Hamiltonian \eqref{eq:genericham} with uncorrelated disorder has a diverging 
mean density of states $\varrho(E) \sim 1/E \log^3(E)$ as $E\rightarrow0$~\cite{Dyson53, Erikhman77, BrouwerPRL00}, 
which also leads to a logarithmic divergence in the localization length with energy~\cite{CohenPRB76, ZimanPRL82}. 
Several independent correlation lengths also diverge for the $E=0$ state~\cite{Fisher94, Fisher95}, indicating 
that the state serves as a disorder induced quantum critical point.
In dynamical properties, the quasilocal nature of the state manifests itself in extremely slow propagation of 
charge~\cite{LuckJSM11} and entanglement growth $\sim \log(\log(t))$~\cite{Fazio06, Vosk13, Sirker16}. 

%%%%%%%%%%%%%%%%%%%%%%%%%%%%%%%%%%%%%%%%%%%%%%%%%%%%%%%%%%%%%%%%%%
% {\it Localization length.}
%%%%%%%%%%%%%%%%%%%%%%%%%%%%%%%%%%%%%%%%%%%%%%%%%%%%%%%%%%%%%%%%%%
We start by investigating the localization length of the model \eqref{eq:genericham} using the transfer matrix 
technique. To this end, we define $\xi_L(E)$ as the localization
length of a finite system of size $L$ at energy $E$. We choose $L$ odd with open boundary condition as it guarantees 
the existence of a $\ez$ state due to sub-lattice symmetry~\footnote{With our system sizes we do not observe any 
even-odd effects in the dynamics. This is because the single particle energy level spacing close to $\ez$ becomes 
exponentially small with $L$}. 
Generically, $\xi_L(E=0)$ can be expressed using the recursion relations between single-particle wavefunction 
amplitudes as, 
\begin{equation}
\xi^{-1}_L(E=0) = \frac{1}{L}\overline{\log\Le\vert\frac{\psi_{L-1}}{\psi_0}\Ri\vert} = 
\frac{1}{L} \sum_{l=1}^{\frac{L-1}{2}} \overline{\log \left (\frac{J_{2l}}{ J_{2l-1}} \right )}, 
\label{eq:xiL}
\end{equation}
where  overline denotes the disorder average. For uncorrelated disorder, e.g., the Dyson I model, 
the average of the summation in Eq.~\eqref{eq:xiL} is zero. However, in a typical configuration the sum is divergent with 
system size $L$, which indicates that one needs to investigate
the full probability distribution of the sequence under the sum rather than just the mean. Using the central limit theorem, 
it can 
be shown that the 
fluctuations grow as $\sqrt{L}$ and therefore $\xi_L(E=0) \sim \sqrt{L}$~\cite{Izrailev12, Che05}. 
On the contrary, in the presence of dimerization, $J_{2l-1}{=}J_{2l}$, the Dyson II model, the sum in 
Eq.~\eqref{eq:xiL} is zero for \emph{each} configuration.
Consequently $\psi_{L-1}=\psi_0$ implying that the $\ez$ state is extended in all samples~\cite{ZimanPRL82}.
With the motivation of interpolating between these two limits of quasilocalized (Dyson I) and extended (Dyson II) $\ez$ states, we choose the random couplings as
\eq{
J_{2l-1} &= \msr{B}_{2l-1}^{(1)}\exp\Le[\frac{-\eta_{2l-1}\msr{B}_{2l-1}^{(2)}}{(2l-1)^\alpha}\Ri];  \nn  \\
J_{2l} &= \msr{B}_{2l-1}^{(1)}\exp\Le[\frac{\eta_{2l} \msr{B}_{2l}^{(2)}}{(2l)^\alpha}\Ri],
\label{eq:distributions}
}
where $\msr{B}_l^{(1)}$, $\msr{B}_l^{(2)}$ are random variables drawn from Gamma distributions with unit mean and variance 
$1/W_{(1,2)}$ defined as 
\eq{
P_{W}(x) = \frac{ W^W}{\Gamma (W) } x^{W-1} e^{-W x};~~~ x \ge 0,
\label{eq:gammadist}
} 
where $\Gamma(W)$ is the Gamma function. 
$\eta_l$'s are independent random variables with the probability density function  $\rho (\eta) =  p \delta(\eta -1) + (1-p) \delta(\eta +1)$ with $p \in [\frac{1}{2},1]$, and $\alpha\ge 0$.  
$J_l$'s are short range correlated random variables and inhomogeneous in space. The inhomogeneity is 
predominantly in the edge of the sample, while in the bulk it is suppressed. 
With this choice of $J_l$'s, Eq.~\eqref{eq:xiL} reduces to
\begin{equation}
\log\Le\vert\frac{\psi_{L-1}}{\psi_0}\Ri\vert = \sum_{l=1}^{L-1}  \frac{ \eta_l \msr{B}_{l}^{(2)}}{l^\alpha}.
\label{eq:xiLbc}
\end{equation} 
In Eq.~\eqref{eq:xiLbc}, $\alpha$ and $p$ determine the asymptotic behavior of $\xi_L(E=0)$ as the thermodynamic limit is approached and also 
allows us to change the extension of the $\ez$ state almost continuously.

\begin{figure}
\includegraphics[width=0.99\columnwidth]{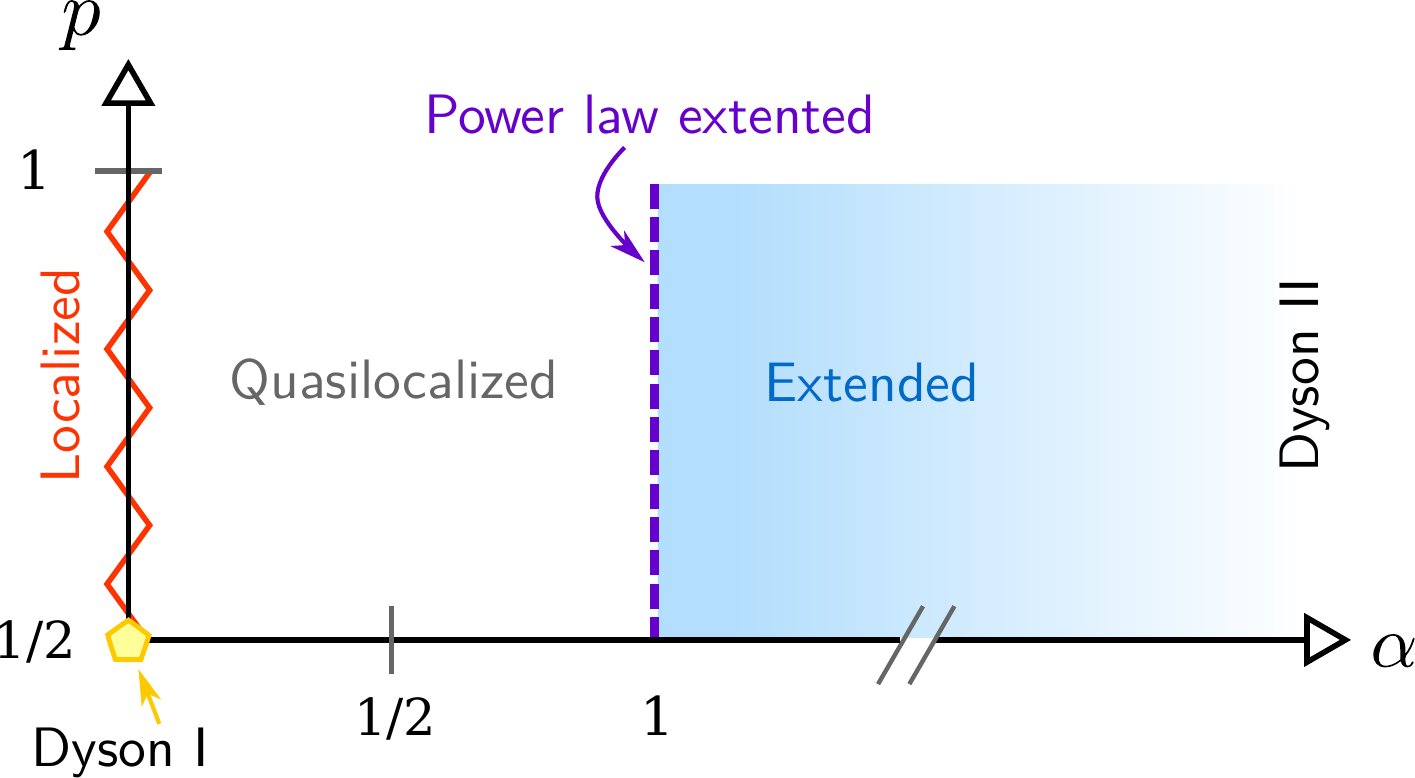}
\caption{Phase diagram with regard to the asymptotic behavior of the $E=0$ state. The regimes denoted by 
`Localized' (`Extended') have localized (extended) $E=0$ state. For $\alpha=0, \, 
p=1/2$ 
limit we recover the  (uncorrelated) Dyson model and also for $\alpha > 1$, the dimerized 
Dyson II model is restored. See text for further details of the localization length in  Eq.~\eqref{eq:xiLpnehalf} and 
Eq.~\eqref{eq:xiLphalf}.}
\label{fig:pd}
\end{figure}

For $p\ne 1/2$ and $\alpha\ge 0$, averaging over the disorder and approximating the sum in Eq.~\eqref{eq:xiLbc} as 
an integral in the large $L$ limit, we get
\eq{
\xi_L(E=0) \sim 
\begin{cases}
(2p-1)^{-1}L^\alpha, & 0\le \alpha<1\\
(2p-1)^{-1}L/\log L, &\alpha=1\\
(2p-1)^{-1}L, &\alpha>1,
\end{cases}
\label{eq:xiLpnehalf}
}
which immediately identifies four distinct regimes. For $\alpha=0$, $\xi_L(E=0)$ is finite, which leads to an 
exponentially localized state. In the range $0<\alpha<1$, the localization length diverges algebraically but slower 
than the system size, which we refer to as a quasilocalized state~(see also Fig.~\ref{fig:pd}). The logarithmic 
correction to $\xi_L(E=0)$ at $\alpha=1$ produces a polynomial spatial decay of the wave function.  
In the limit $\alpha\rightarrow \infty$, the correlation reveals itself via the 
dimerization of bonds, $J_{2l-1} = J_{2l}$, which is the Dyson II model with an extended $\ez$ state.  
%%%%%%%%%%%%%%%%%%%%%%%%%%%%%%%%%%%%%%%%%%%%%%%%%%%%%%%%%%%%%%%  
%%%%%%%%%%%%%%%%%%%%%%%%%%%%%%%%%%%%%%%%%%%%%%%%%%%%%%%%%%%%%%%
%%%%%%%%%%%%%%%%%%%%%%%%%%%%%%%%%%%%%%%%%%%%%%%%%%%%%%%%%%%%%%%%%%%%%%%%%%%%%%%%%%
\begin{figure*}[t]
 \includegraphics[width=0.975\textwidth]{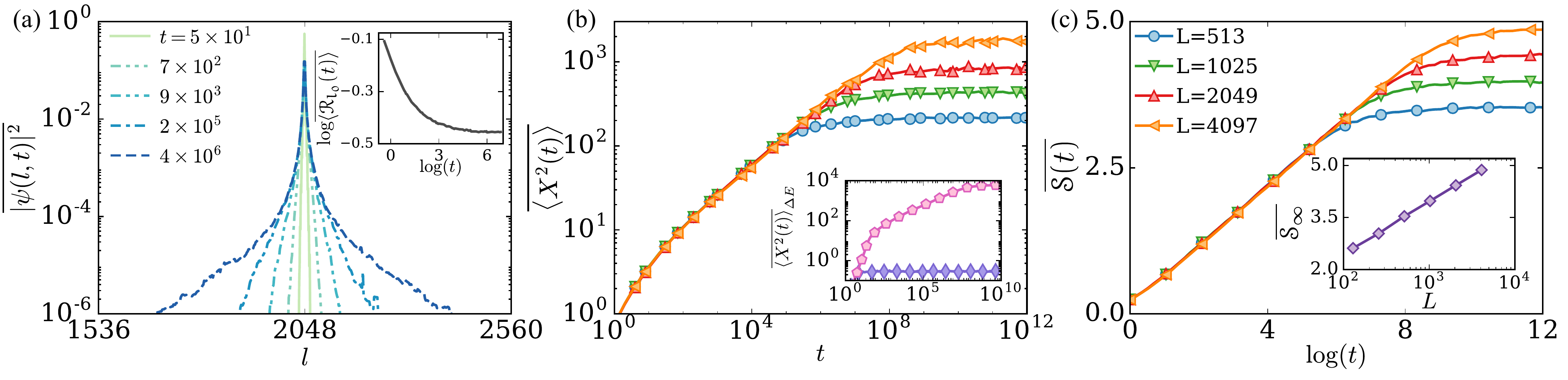}
 \caption{
 (a) The disorder averaged wavepacket at different times for the Dyson II model. The central core decays quickly and 
saturates after initial dynamics; whereas the tail of the distribution keeps spreading with time. Inset shows 
the return probability for $L=4097$. 
 (b) The growth of $\overline{\langle X^2(t)\rangle}$ with time for $L=\{513, \ldots, 4097\}$ in log-log scale. For 
finite systems 
it saturates to a value which grows linearly with the 
system size. Inset shows $\overline{\langle X^2(t)\rangle}_{\Delta E}$ with $\ez$ present(\pentagon) in $\Delta E$ which 
grows subdiffusively  and absent ($\lozenge$) which saturates, hence confirming that the dynamics is governed by 
the states close to $E=0$~($L=4097$). (c) The entanglement entropy shows a logarithmic growth in time
 $\overline{\msr{S}(t)} \sim \log t$ and the saturation, $\overline{\msr{S}_\infty}$, 
  grows logarithmically with $L$ as shown in the inset.}
 \label{fig:type2}
\end{figure*}
%%%%%%%%%%%%%%%%%%%%%%%%%%%%%%%%%%%%%%%%%%%%%%%%%%%%%%%%%%%%%%%%%%%%%%%%%%%%%%%%%%

For $p=1/2$, the sign $\eta_l$ appears with equal probability. Therefore, $\xi^{-1}_L(E=0)$ defined in 
Eq.~\eqref{eq:xiL} goes to zero upon taking disorder average. Hence, in order to
understand the asymptotic behavior of $\xi^{-1}_L(E=0)$, we analyze the
fluctuations of the sequence $\{\log\Le\vert{\psi_{L-1}}/{\psi_0}\Ri\vert \}$, similar to the Dyson I model as follows.
Let $\msr{A}_L$ be the random variable defined after averaging over $\msr{B}_l^{(2)}$s in Eq.~\eqref{eq:xiLbc}, 
\eq{
 \msr{A}_L = \overline{\Le[\log\Le\vert\frac{\psi_{L-1}}{\psi_0}\Ri\vert\Ri]}
=\sum_{l=1}^{L-1}\frac{\eta_l }{l^\alpha}.
\label{eq:Al}
 }
$\msr{A}_L$ is a sum of independent but not identically
distributed random variables with zero mean and variance $\sigma^2_l=1/l^{2\alpha}$.
The \emph{Lyapunov Central Limit 
theorem}~\cite{koralov2007theory} then dictates that 
the probability distribution of $\msr{A}_L$ approaches to a Gaussian
distribution with zero mean and variance, $\sigma^2_{\msr{A}_L}=\sum_{l=1}^{L-1}l^{-2\alpha}$, in the limit 
$L\rightarrow \infty$. The asymptotic behavior of $\sigma^2_{\msr{A}_L}$ can then be used to extract the behavior of 
the localization length,  
\eq{
\xi_L(E=0) \propto \f{L}{\sigma_{\msr{A}_L}} \sim 
\begin{cases}
L^{\alpha + 1/2}, &0\le\alpha<1/2\\
L/\sqrt{\log L}, &\alpha=1/2\\
L, &\alpha>1/2.
\end{cases}
\label{eq:xiLphalf}
}
Three qualitatively different regimes can be identified. For $0\le\alpha<1/2$,
the localization length diverges algebraically, but slower than the system size.
At the $\alpha=0,\,p=1/2$ point, we recover the Dyson I model, where the localization length diverges as $\sim 
\sqrt{L}$ solely due to fluctuations. Finally, for $\alpha>1/2$, the state is extended with system size. 
The behavior of $\xi_L(E=0)$ as a function of $\alpha$ and $p$ is summarized in Fig.~\ref{fig:pd}. Importantly, the 
phase diagram is stable against any local perturbations that do not break the original 
symmetry of the $\mc{H}$, because it does not qualitatively change the structure of Eq.~\eqref{eq:xiL}.

%%%%%%%%%%%%%%%%%%%%%%%%%%%%%%%%%%%%%%%%%%%%%%%%%%%%%%%%%%%%%%%%%%%%%%%%%%%%%%%%%
% {\it Dynamical properties.}
\section{Dynamical properties \label{sec:dynamical}}
%%%%%%%%%%%%%%%%%%%%%%%%%%%%%%%%%%%%%%%%%%%%%%%%%%%%%%%%%%%%%%%%%%%%%%%%%%%%%%%%%
Having established that the model \eqref{eq:genericham} with the random couplings  
\eqref{eq:distributions} hosts several different natures of extended/quasilocalized state at $E=0$, we now 
investigate its effects on dynamical properties. 
First, we study charge propagation via wavepacket dynamics in the single particle framework~\cite{Iz97, 
Ketzmerick97, Huckestein99, Nazareno05}.
The initial wavepacket is localized at a single point $l_0$ in the middle of the 
chain, $\psi(l,t=0)=\delta_{l,l_0}$. 
With time it spreads out and its amplitude at the initial site $l_0$ decays. We monitor the decay of 
the initial density via the return probability $\la \msr{R}_{l_{0}}(t)\ra = 
|\psi(l_0, 
t)|^2$ and quantify the spreading of the charge by the disordered average mean-square displacement
$
\overline{\Le\langle X^2(t)\Ri\rangle}=\sum_{l} l^2 \overline{\vert\psi(l,t)\vert^2} - (\sum_{l} l 
\overline{\vert\psi(l,t)\vert^2})^2. 
$
Furthermore, the growth of bipartite entanglement entropy 
$
\msr{S}(t) {=} -\text{Tr} (\rho_\text{L}(t) \log(\rho_\text{L}(t)))
$, between two halves of the system L and R is investigated using standard free fermion 
techniques~\cite{Peschel03}, where $\rho_\text{L}(t) {=} 
\text{Tr}_\text{R}(|\psi(t) \ra \la \psi(t)|)$ 
and $|\psi(t{=}0)\ra$ is a random product state at half-filling. 
Under time-evolution, L and R subsystems exchange information leading
to the growth of $\msr{S}(t)$, which is zero at $t=0$. 
In our simulations, we use open boundary conditions with $W_1=0.4$ and $W_2=10$, and checked (not shown), that 
with periodic boundary condition,  even number of sites and also with other values of $W_{(1,2)}$ there are no 
qualitative difference in the conclusions.

%%%%%%%%%%%%%%%%%%%%%%%%%%%%%%%%%%%%%%%%%%%%%%%%%%%%%%%%%%%%%
% {\it Results, Dyson II, ($\alpha{\rightarrow}\infty$).}
\subsection{Dyson II ($\alpha{\rightarrow}\infty$)  \label{sec:dyson2}}
%%%%%%%%%%%%%%%%%%%%%%%%%%%%%%%%%%%%%%%%%%%%%%%%%%%%%%%%%%%%%
Since the dynamical properties of these localized systems is expected to be
dominated by the properties of the states close to $\ez$, it is expected that the 
dynamics would be qualitatively different depending on which
regime of the phase diagram they belong to.  We first focus on the Dyson II model with dimerized hopping. In 
Fig.~\ref{fig:type2}(a) we show the probability distribution of the time dependent wavefunction at different times. At 
long times only the tail of the wavefunction keeps spreading, while the return probability saturates after an 
algebraic decay as seen in the inset. Finite  $\overline{\la \msr{R}_{l_0}( t)\ra}$ at long times implies a finite 
density of exponentially localized states in the energy spectrum.~\footnote{Using numerical transfer matrix 
calculation, we checked that at other energies~($E\ne 0$) corresponding Lyapunov exponents are strictly positive.}

Fig.~\ref{fig:type2}(b) shows the expansion of the width of wavepacket. 
The linear behavior of the width with time in log-log scale suggests
$\overline{\langle X^2(t)\rangle}\sim t^{\beta}$, where the non-universal exponent
$\beta$ depends on the disorder strength, e.g, $\beta\approx0.35$ for $W_{1}=0.4$ , which implies subdiffusion. 
For finite systems, the growth saturates, with the saturation value growing linearly with the system size 
reflecting the spatial extension of the $\ez$ state~\eqref{eq:xiLpnehalf}.  

Note that due to the diverging nature of the density of states, the dynamics is always going to be dominated 
by a finite number of states in the vicinity of $\ez$. 
We ascertain this by projecting the initial 
wavepacket onto eigenstates within an energy window $\Delta E$ that includes $\ez$ and also away from it as 
$\vert\psi_0\rangle_{\Delta E}=\hat P_{\Delta E}\vert\psi(l_0,t{=}0)\rangle$, where $\hat{P}_{\Delta E}=\sum_{E\in\Delta 
E} |E\ra \la E|$ and $|E\ra$ is the eigenstate. We contrast the two situations by 
measuring the spread of the wavepacket as $\overline{\langle X^2(t)\rangle}_{\Delta 
E}=\overline{\langle\psi_0\vert_{\Delta E} \hat{X}^2(t)\vert\psi_0\rangle_{\Delta 
E}}-\overline{\langle\psi_0\vert_{\Delta E} \hat{X}^2(0)\vert\psi_0\rangle_{\Delta E}}$.
As seen in Fig.~\ref{fig:type2}(b, inset) the spectral 
decomposed wavepacket with the $E=0$ state shows a subdiffusive propagation~(\pentagon), whereas the 
wavepacket that has been projected away from the band center quickly saturates~($\lozenge$) as one would expect for 
localized states.

Fig.~\ref{fig:type2}(c) shows the growth of disorder averaged bipartite entanglement $\overline{\msr{S}(t)}$ starting 
from a product state. We observe a logarithmic growth of $\overline{\msr{S}(t)}$ in time,  
which is slower than the charge transport. For $W_1=0.4$ the prefactor of $\log(t)$ is $\approx \ln(2)/3$. 
In the inset of Fig.~\ref{fig:type2}(c) the 
saturation value of $\overline{\msr{S}(t)}$ at $t\rightarrow\infty$  ($\overline{\msr{S}_\infty}$) is plotted in a log-linear scale, which shows 
logarithmic scaling with system size with a slope $\approx \ln(2)$. The 
logarithmic scaling of $\overline{\msr{S}_\infty}$ is similar to entanglement scaling of critical states. 
Unlike in an interacting localized phase, where entanglement is generated via dephasing due to 
interaction~\cite{Bardarson12, AbaG13}, here it is due to the extended nature of the $\ez$ state, which implies that 
the saturation time of $\msr{S}(t)$ is proportional to the localization length of the extended state. 

Note that, there is no qualitative change in our results at higher values of $W_1$. Specifically for $W_1>1$, when the Gamma distribution \eqref{eq:gammadist} becomes non-singular at zero, $\overline{\langle X^2(t) \rangle}$ and $\overline{\msr{S}(t)}$ still show a subdiffusive and logarithmic growth in time respectively, as shown in Appendix~\ref{sec:app}.
%%%%%%%%%%%%%%%%%%%%%%%%%%%%%%%%%%%%%%%%%%%%%%%%%%%%%%%
\begin{figure}[tb]
\includegraphics[width=1.0\columnwidth]{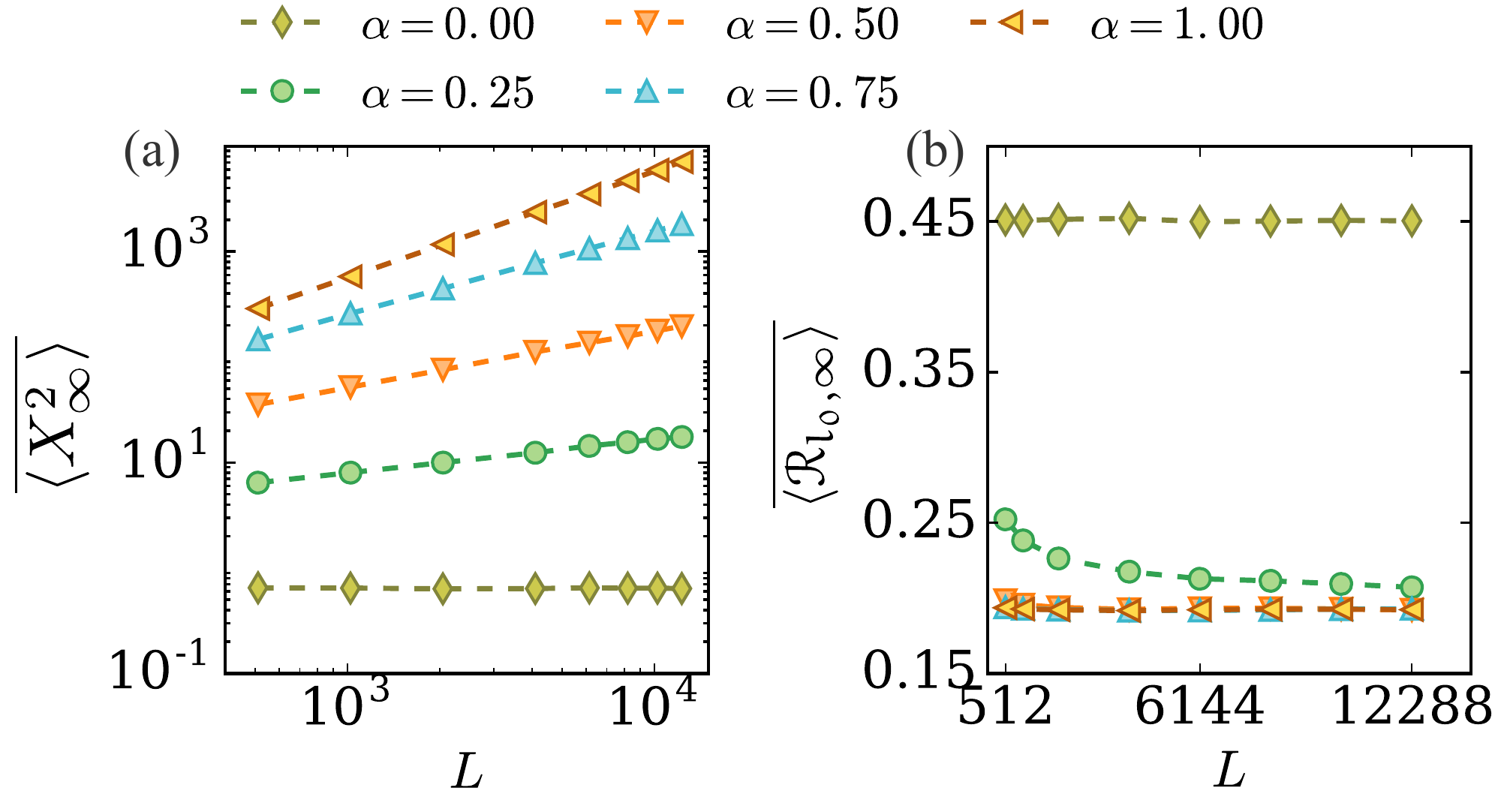}
\caption{(a) $\overline{\la X^2_\infty\ra}$ for different values of $\alpha$ in a log-log scale to 
highlight the scaling $\propto L^\alpha$ as expected from the localization length calculation~\eqref{eq:xiLpnehalf}. 
(b) The return probability $\overline{\la\msr{R}_{l_0, \infty}\ra}$~\eqref{eq:ipr} for different $\alpha$ shows 
saturation with system size $L$. Dashed lines are given as guides to the eye.}
\label{fig:X2_R_type3}
\end{figure}
%%%%%%%%%%%%%%%%%%%%%%%%%%%%%%%%%%%%%%%%%%%%%%%%%%%%%%%
% {\it  Results, $ 0 \le \alpha\le 1, \,p{=}1$.} 
\subsection{ $ 0 \le \alpha\le 1, \,p = 1$ \label{sec:dyson3}} 
%%%%%%%%%%%%%%%%%%%%%%%%%%%%%%%%%%%%%%%%%%%%%%%%%%%%%%%
For any finite $\alpha$, charge propagation is subdiffusive. 
The difference for different $\alpha$ is seen in the scaling of the saturation values of $\overline{\la 
\msr{R}_{l_0, \infty} \ra}$ and $\overline{\la 
X^2_\infty\ra} $ with $L$, as the localization lengths depend on $\alpha$. 
Fig.~\ref{fig:X2_R_type3}(a) shows the $t\rightarrow\infty$ value of the width of the wavepacket in a 
log-log plot as a function of system size. The  
leading behavior is given by $L^\alpha$ as one would expect from the extended nature of the $E=0$ 
eigenstate described in Eq.~\eqref{eq:xiLpnehalf}. Crudely approximating the $E=0$ eigenstate, $\vert\phi_0\rangle$, as a box-function of width $\xi_L(E=0)$, one finds $\overline{\langle\phi_0\vert\hat{X}^2\vert\phi_0\rangle}\propto\xi_L(E=0)$.
Similarly, in Fig.~\ref{fig:X2_R_type3}(b) we show the return probability at 
$t\rightarrow\infty$, defined 
as
\eq{
\la \msr{R}_{l_0}(t)\ra {=}  \Le|\psi(l_0,t)\Ri|^2 \xrightarrow[]{t=\infty}  \sum_n 
\Le|\phi_n(l_0)\Ri|^4, 
\label{eq:ipr}
}
which is the inverse participation ratio of the single particle eigenstates. 
Two things are of note: (i) for $0 \le \alpha \le 1$, it converges with $L$, which emphasizes that 
most of the eigenstates are localized, (ii) for $\alpha=0$, the $\overline{\la \msr{R}_{l_0, \infty} \ra}$  
converges at a different value than other $\alpha$'s. 
This can be understood from the following decomposition of inverse participation ratio~\eqref{eq:ipr}, 
$
\sum_n \Le|\phi_n(l_0)\Ri|^4 
= \sum_{n < |\Delta E|} \Le|\phi_n(l_0)\Ri|^4 + \sum_{n> |\Delta E|} \Le|\phi_n(l_0)\Ri|^4
$
, where $\Delta E$ is 
the window of energies enclosing delocalized states around $E=0$. Only for $\alpha \ne 0$ the first term in the sum is 
negligible because of the extended nature of the states within the interval $\Delta E$, however for $\alpha=0$, 
$\Delta E=0$ as all states are localized~\eqref{eq:xiLpnehalf}. Therefore, it is 
expected that $\alpha=0$ converges at a higher value as seen in Fig.~\ref{fig:X2_R_type3}(b) compared to other 
$\alpha$.

%%%%%%%%%%%%%%%%%%%%%%%%%%%%%%%%%%%%%%%%%%%%%%%%%%%%%%%
\begin{figure}[tb]
\includegraphics[width=0.95\columnwidth]{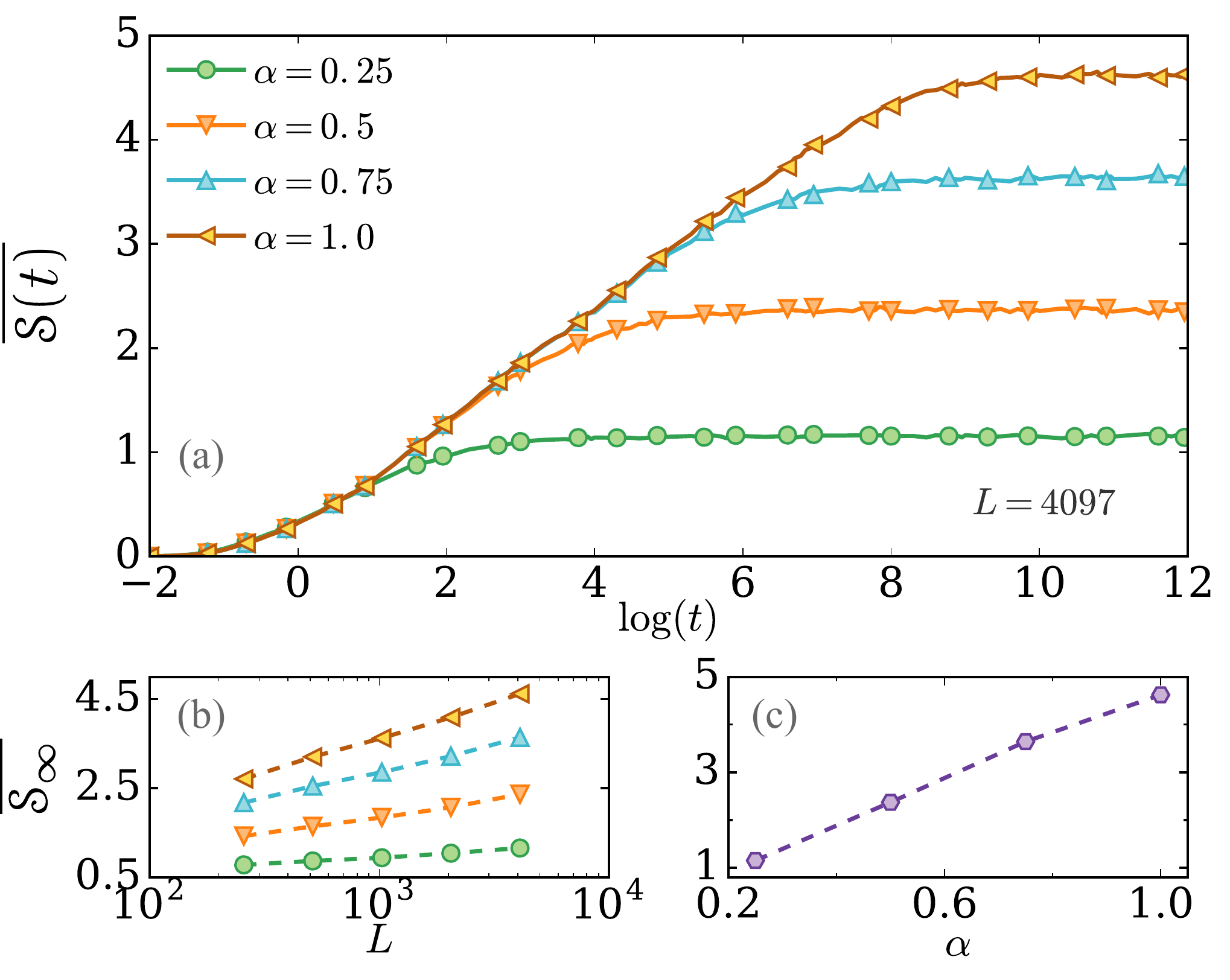}
\caption{(a) Dynamics of entanglement for different values of $\alpha$ and $p=1$ in a log-linear scale 
after a quench from a product state. The logarithmic growth of $\overline{\msr{S}(t)}$ is visible for 
all values of $\alpha$ shown here. (b) The saturation value of $\overline{\msr{S}(t)}$ at long time behaves as 
$\log(L)$ for all $\alpha\ne0$. (c) The entanglement saturation $\overline{\msr{S}_\infty}$ shows a linear growth 
with $\alpha$ ($L=4097$)~\eqref{eq:entg_law}. }
\label{fig:Ent_type3}
\end{figure}
%%%%%%%%%%%%%%%%%%%%%%%%%%%%%%%%%%%%%%%%%%%%%%%%%%%%%%%
%%%%%%%%%%%%%%%%%%%%%%%%%%%%%%%%%%%%%%%%%%%%%%%%%%%%%%%
Fig.~\ref{fig:Ent_type3}(a) shows the time evolution of $\overline{\msr{S}(t)}$ for 
different values of $\alpha$ after a global quench. The data shows a logarithmic growth of entanglement similar to 
Dyson II. Note that the slope at which $\msr{S}(t)$ grows is almost independent of $\alpha$, while the effect 
of $\alpha$ is clearly visible in the saturation. To highlight the dependence of the saturation 
with system size we plot $\overline{\msr{S}_\infty}$ as a function of $L$ in 
Fig.~\ref{fig:Ent_type3}(b) in log-linear scale. For $\alpha > 0$ we see a logarithmic increase of  
$\overline{\msr{S}_\infty}$ with a slope $\propto \alpha$. This is further confirmed in 
Fig.~\ref{fig:Ent_type3}(c), where the saturation of entanglement is plotted as a function of $\alpha$. The behavior 
suggests the following form of $\msr{S}(t)$ with time and system size,  
\eq{
\overline{\msr{S}(t)} &\sim \log(t); \quad \; \overline{\msr{S}_\infty} \sim \log[\xi_{L, \alpha}(E=0)] 
\label{eq:entg_law}
}
where $\xi_{L, \alpha}(E=0)$ is the localization length and is $\propto L^\alpha$~\eqref{eq:xiLpnehalf}. For 
$\alpha=0, \, p>1/2$ the state is exponentially localized and therefore neither charge or entanglement 
propagate. 

%%%%%%%%%%%%%%%%%%%%%%%%%%%%%%%%%%%%%%%%%%%%%%%%%%%%%%%
% {\it Discussions.}
\section{Conclusion \label{sec:conclusion}}
%%%%%%%%%%%%%%%%%%%%%%%%%%%%%%%%%%%%%%%%%%%%%%%%%%%%%%%
In summary, we have constructed a generalized correlated 
one-dimensional random bond disorder model and studied its non-equilibrium dynamics. Even though the localization 
length of the $\ez$ state is divergent, the state can be quasilocalized or extended and its spatial extent depends on the 
correlations in disorder. We have shown that the dynamical properties are dominated by the  states close to 
$\ez$. In all the parameter regimes studied we find subdiffusive transport, while logarithmically slow 
growth of entanglement. The saturation value of the wavepacket and entanglement depends on the finite size 
localization length of the $\ez$ state. In particular, $\overline{\msr{S}_{\infty}}$ grows logarithmically with 
the localization length of the $\ez$ state. The scaling behavior is similar to the scaling of $\msr{S}$ in the excited 
state of uncorrelated random spin chain in the same universality class~\cite{HuangPRB14,Vasseur15}, except that in our 
generalized model disorder correlation enters in the $\overline{\msr{S}_{\infty}}$ scaling via the finite size 
localization length of the $\ez$ state.

\begin{acknowledgments}
We thank A. B\"acker, D. Bagrets, A. Croy, F. Evers, A. Lazarides, R. Singh, 
and J-M. St\'ephan for several illuminating discussions. We also express our gratitude to J. H. Bardarson, F. 
Evers, and F. Pollmann for a critical reading of the manuscript. 
\end{acknowledgments}

\begin{appendix}
%%%%%%%%%%%%%%%%%%%%%%%%%%%%%%%%%%%%%%%%%%%%%%%%%%%%%55
\section{Results for different disorder strengths \label{sec:app}}
%%%%%%%%%%%%%%%%%%%%%%%%%%%%%%%%%%%%%%%%%%%%%%%%%%%%%%
\begin{figure}[!t]
\includegraphics[width=1\columnwidth]{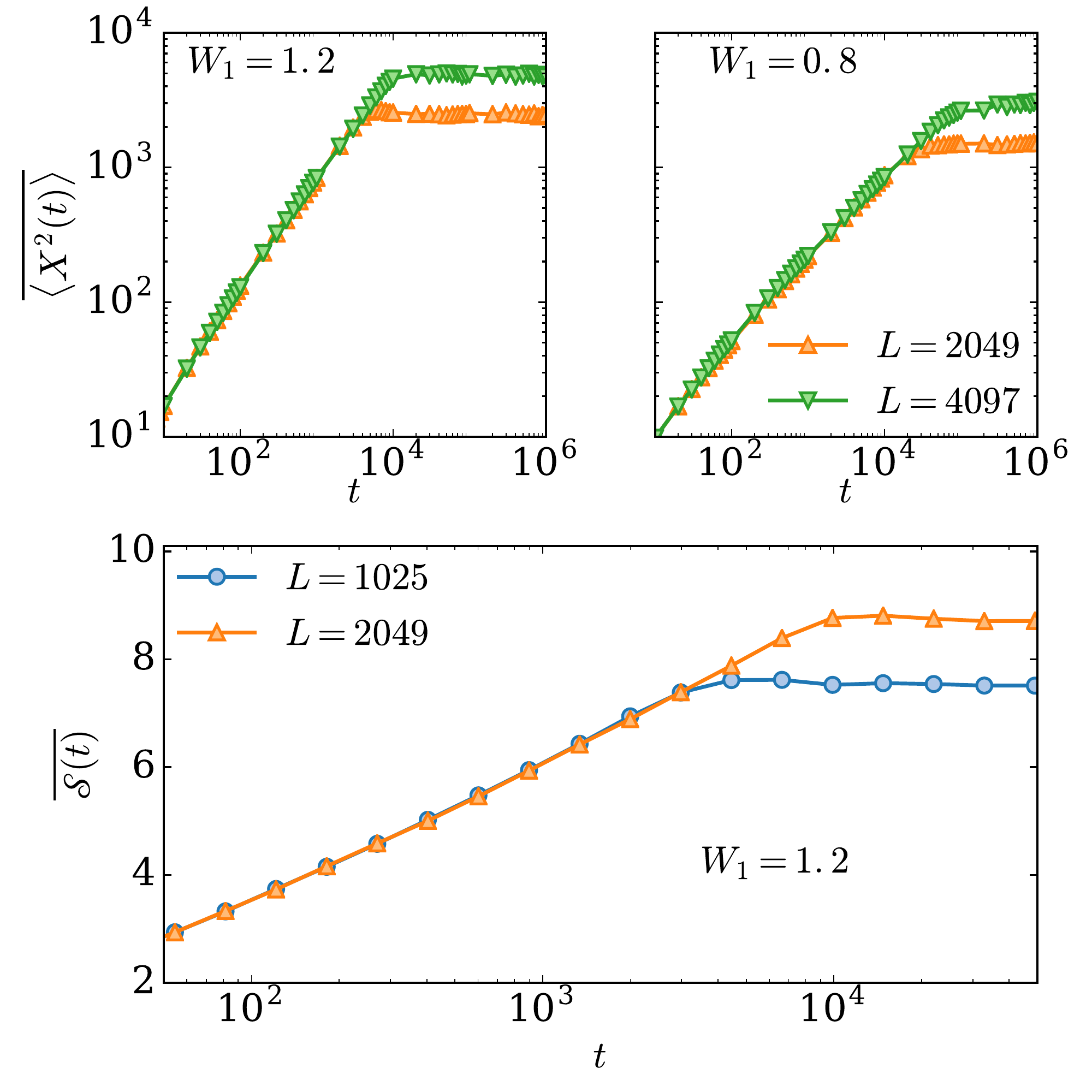}
\caption{(Top) The subdiffusive growth of $\overline{\langle X^2(t)\rangle}$ for the Dyson II model in log-log scale for two different values of $W_1$ and for two different system sizes $L = 2049, 4097$. (Bottom) Logarithmic growth of $\overline{\mathscr{S}(t)}$ for the Dyson II model for $W_1 = 1.2$ and $L = 1025, 2049$.}
\label{fig:x2_diffW}
\end{figure}
In this appendix we show additional results for different values of $W_1$ for the Dyson II model. They further substantiate our conclusions about subdiffusive wavepacket dynamics, and logarithmically slow entanglement growth in the generalized model. 
Fig.~\ref{fig:x2_diffW} (top) shows the growth of $\overline{\langle X^2(t)\rangle}$ for the Dyson II model for $W_1 = 0.8$ and $W_1= 1.2$.
For both these values of $W_1$, $\overline{\langle X^2(t)\rangle}$ grows algebraically with time $ ( \overline{\langle X^2(t)\rangle} \sim  t^{\beta(W_1)}$ , with  $\beta( W_1 = 1.2) \approx 0.78$ and   $\beta( W_1 = 0.8) \approx 0.59$, 
showing the subdiffusive dynamics.  
Fig.~\ref{fig:x2_diffW} (bottom) shows that the growth of $\overline{\msr{S}(t)}$ for the Dyson II model with $W_1 = 1.2$. It is still clearly visible that the entanglement growth in time is logarithmic, $\overline{\mathscr{S}(t)} \sim \log(t)$. 
Note that, for $W_1=1.2$, the Gamma distribution is no longer singular at zero, yet we see subdiffusive wavepacket dynamics and logarithmic entanglement growth, ensuring that this behavior is indeed generic.

 %%%%%%%%%%%%%%%%%%%%%%%%%%%%%%%%%%%%%%%%%%%%%%%%%%%%%%%

%%%%%%%%%%%%%%%%%%%%%%%%%%%%%%%%%%%%%%%%%%%%%%%%%%%%%%%

%%%

\end{appendix}

%%%%%%%%%%%%%%%%%%%%%%%%%%%%%%%%%%%%%%%%%%%%%%%%%%%%%%%

%%%%%%%%%%%%%%%%%%%%%%%%%%%%%%%%%%%%%%%%%%%%%%%%%%%%%%%

%%%%%%%%%%%%%%%%%%%%%%%%%%%%%%%%%%%%%%%%%%%%%%%%%%%%%%%%%5
\bibliography{ref}
%%%%%%%%%%%%%%%%%%%%%%%%%%%%%%%%%%%%%%%%%%%%%%%%%%%%%%%%%5
\end{document}